\documentclass{iau}
\usepackage{graphicx,natbib}
 
\newcommand{\atlas}{{{ATLAS$^{\mathrm{3D}}$}}}
\newcommand{\afe}{[$\alpha$/Fe]}

\title{Connection between dynamically derived IMF normalisation and stellar populations}

\author[R.M. McDermid]{Richard M. McDermid$^{1,2}$}

\affiliation{$^{1}$Department of Physics and Astronomy, Macquarie University, Sydney NSW 2109, Australia\\
$^{2}$Australian Astronomical Observatory, PO Box 915, Sydney NSW 1670, Australia\\
email: {\tt richard.mcdermid@mq.edu.au}\\
}

\pubyear{2014}
\volume{311}
\jname{Galaxy Masses as Constraints of Formation Models}
\editors{M. Cappellari \& S. Courteau, eds.}

\begin{document}

\maketitle

\begin{abstract}
In this contributed talk I present recent results on the connection between stellar population properties and the normalisation of the stellar initial mass function (IMF) measured using stellar dynamics, based on a large sample of 260 early-type galaxies observed as part of the \atlas\/ project. This measure of the IMF normalisation is found to vary non-uniformly with age- and metallicity-sensitive absorption line strengths. Applying single stellar population models, there are weak but measurable trends of the IMF with age and abundance ratio. Accounting for the dependence of stellar population parameters on velocity dispersion  effectively removes these trends, but subsequently introduces a trend with metallicity, such that `heavy' IMFs favour lower metallicities. The correlations are weaker than those found from previous studies directly detecting low-mass stars, suggesting some degree of tension between the different approaches of measuring the IMF. Resolving these discrepancies will be the focus of future work.\looseness-2


\end{abstract}

\firstsection
\section{Introduction}

A universal form for the stellar initial mass function (IMF) has been an actively debated topic for many decades, having far-reaching consequences for our understanding of the star formation process, and more generally, of how to interpret the integrated light we observe coming from unresolved stellar populations in galaxies and star clusters. On one hand, the environment (i.e. physical conditions) for star formation and the evolutionary processes at work can vary significantly from galaxy to galaxy, suggesting that a non-universal IMF should not be ruled out. On the other, until recently there was little compelling evidence that could sufficiently isolate the properties of the IMF in such unresolved systems, and thus constrain its approximate functional form.

A recent breakthrough was the convincing detection by \citet{vdc10} of an over-abundance of low-mass stars compared to fiducial IMFs. Subsequently, strong evidence for a non-universal IMF has emerged from the application of various independent techniques, including gravitational lensing \citep{auger10}, stellar dynamical modelling \citep{cappellari12, cappellari13b, tortora13, conroy13}, and spectral synthesis \citep{vdc10, vdc11, spiniello12, labarbera13, ferreras13, spiniello14}.

Here we use the IMF derived via the stellar dynamical modelling of \citet{cappellari12} to explore whether the IMF - a key property of any stellar population - shows any measurable systematic variations with galaxy stellar population parameters, as derived via line-strengths and single stellar population models. This summarises work presented in \citet{mcdermid14} and \citet{mcdermid15}.

\begin{figure}
\centering
\includegraphics[width=0.4\columnwidth]{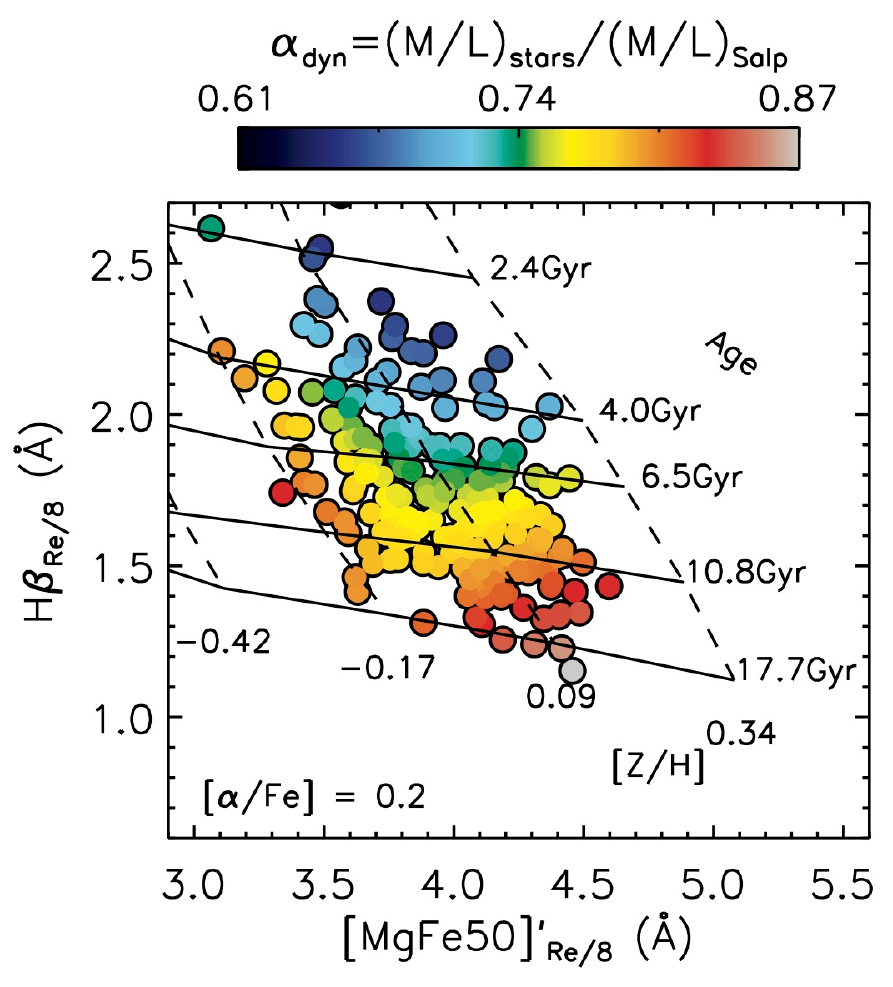} 
\caption{Lick index H$\beta$ versus the combined index $[$MgFe50$]^\prime$, both measured within an aperture of one eighth of an effective radius. The colour scale indicates the values of the IMF parameter, $\alpha_{\rm dyn}$, which have been adaptively smoothed using a locally-weighted regression technique to show underlying trends in the distribution. A grid of SSP model predictions from \cite{schiavon07} is shown for a super-solar abundance ratio of \afe\/$=0.2$, as indicated in the lower left of the plot. Solid lines indicate lines of constant age; dashed lines show constant metallicity.}
\label{fig:fig1}
\end{figure}

\section{Methods}

We quantify the mass normalisation of the IMF using the IMF parameter $\alpha_{\rm dyn}\equiv(M/L)_{\rm stars}/(M/L)_{\rm Salp}$, such that $\alpha_{\rm dyn}>1$ implies an IMF `heavier' than the fiducial unimodal power law of the form: $\zeta(m) \propto m^{-2.35}$ \citep{salpeter55}.  The nominator of this ratio is derived from fitting dynamical models to the observed integral-field stellar kinematics. These models account for differing orbital anisotropy between galaxies, and implicitly reproduce the shape and profile of the observed surface brightness to high accuracy. Additionally, the presence of dark matter was explored using various general prescriptions for the halo mass profile, including contracted haloes. In this way, the inferred mass-to-light ratio entering into the $\alpha_{\rm dyn}$ parameter is dynamical measure of the stars {\it only}. The dynamical models are described in full in \cite{cappellari13a}.

The denominator of the $\alpha_{\rm dyn}$ parameter is derived from fitting stellar population models to the \atlas\/ optical spectroscopy, as described in \citet{mcdermid15}. Using the penalised pixel fitting method of \citet{cappellari04}, the spectra are fitted with a (positive) regularised linear combination of the stellar population models of \citet{vazdekis12}, fitting for age and metallicity as free parameters, and holding the IMF of the models fixed at Salpeter. With this approach, all the information in the spectrum is used to constrain the spread in possible ages and metallicities, and thus accounts for variations in star formation history between different galaxies.

Finally, the stellar population parameters are measured using line strengths and the single stellar population (SSP) models of \citet{schiavon07}. These models have variable abundance ratios, encapsulated in the ratio of $\alpha$-elements to iron, \afe, in addition to age and total metallicity, [Z/H]. The measurements are taken from \citet{mcdermid15}, from which we use the central apertures covering $R_e/8$ to be consistent with existing spectral IMF studies.

\vspace{-0.5cm}
\section{Results}

Fig.~\ref{fig:fig1} presents the averaged distribution of $\alpha_{\rm dyn}$ values on the plane of H$\beta$ versus the combined index $[$MgFe50$]^\prime$ of \citet{kuntschner10}, over plotted with the Schiavon model grid for \afe\/$=0.2$. There is a trend of larger values of the IMF parameter (corresponding to `heavier' IMFs) towards weaker H$\beta$ (older), and stronger $[$MgFe50$]^\prime$ (metal-rich) objects. The trend, however, is not monotonic, with lower metallicity objects also showing relatively high IMF parameter values, on average.

\begin{figure}
\centering
\includegraphics[width=0.477\columnwidth]{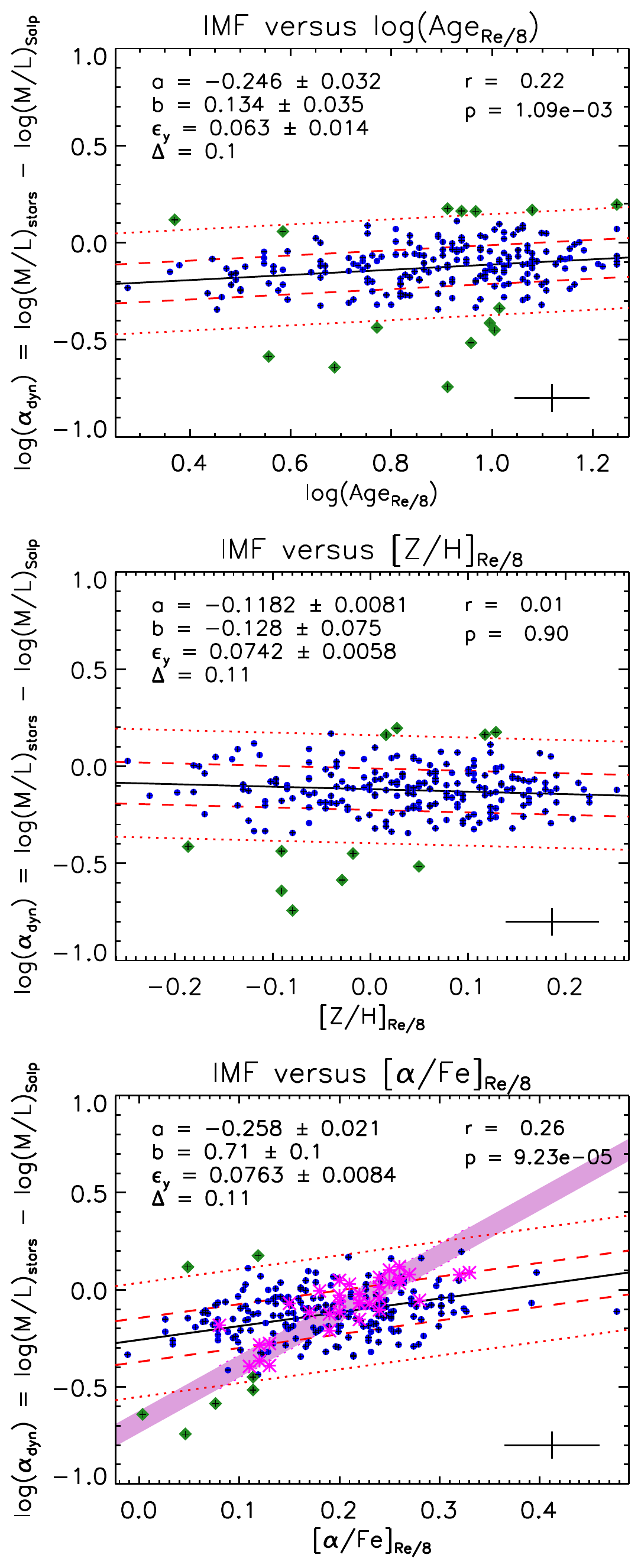} 
\includegraphics[width=0.51\columnwidth]{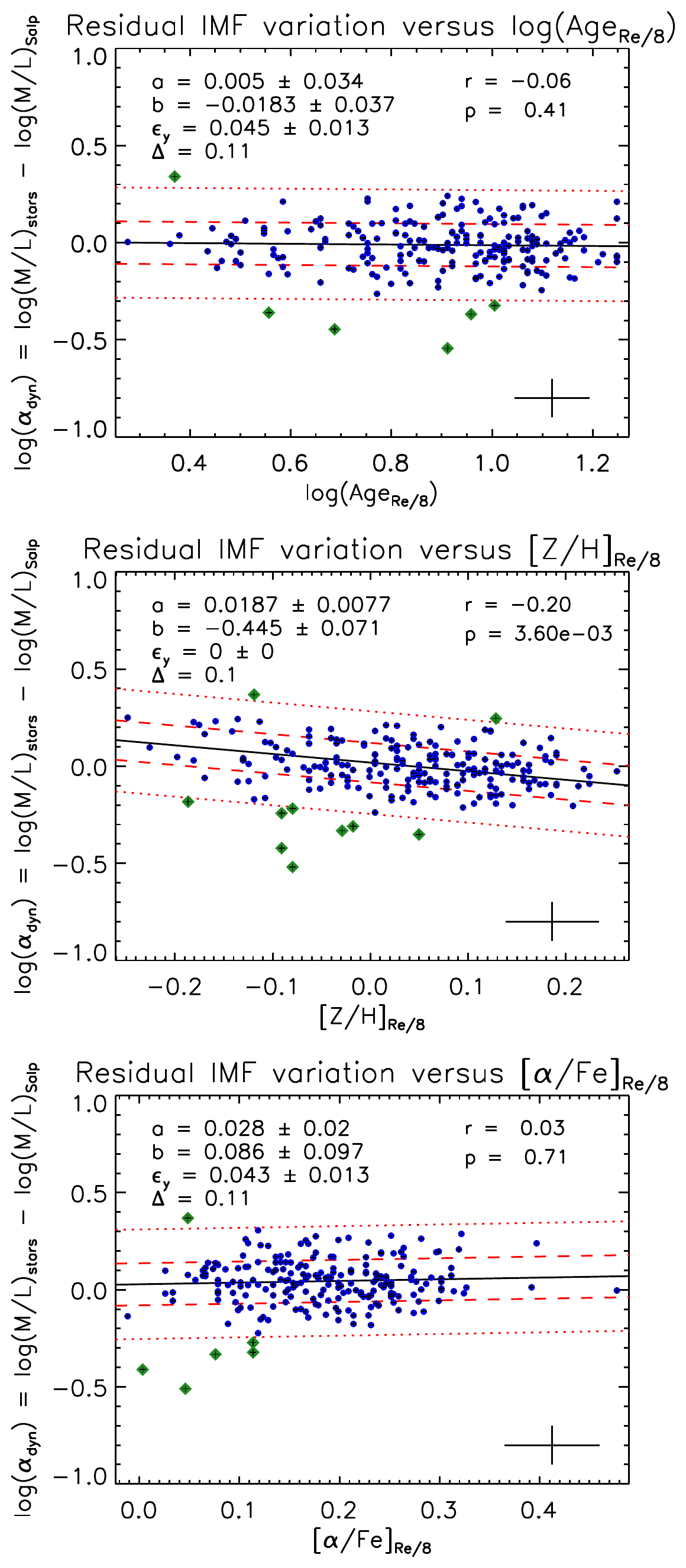} 
\caption{{\it Left:} Relations of IMF parameter (y-axis) with (top to bottom) age, total metallicity and \afe, showing in each panel the linear fit parameters (upper left) and correlation coefficient (r) and significance (p, upper right). The bottom panel also shows points from \citet{cvd12} in magenta, over-plotted with a band indicating the linear fit obtained using the same algorithm as for the \atlas\/ points. {\it Right:} Equivalent plots to left column, but after the co-dependence on velocity dispersion has been removed. This removes the trends with age and \afe, but a trend with metallicity becomes apparent.}
\label{fig:fig2}
\end{figure}

The left panels of Fig.~\ref{fig:fig2} presents the trends of $\alpha_{\rm dyn}$ with the three measured SSP parameters, together with robust linear fits, fit parameters, correlation coefficients (r), and correlation significance (p). These show weak but significant trends of $\alpha_{\rm dyn}$ with age and \afe, and no measured correlation with [Z/H]. For reference, we over-plot the data points of \citet{cvd12} in magenta, together with a region showing the slope and 1$\sigma$ (68\%) scatter derived from applying the same line-fitting procedure to these points. The relation from the \atlas\/ points is notably less steep.

Stellar population parameters are known to positively correlate with velocity dispersion \citep{thomas05,graves09,mcdermid15}, as does $\alpha_{\rm dyn}$ \citep{cappellari13b}. \citet{smith14} has shown that accounting for this additional dependency can change the interpretation of what primarily drives the inferred IMF variations. To explore this, the right-hand panels of Fig.~\ref{fig:fig2} show the residual correlations after removing the dependency of the stellar population parameters on velocity dispersion. This is done by fitting a plane to the three-dimensional distribution of $\alpha_{\rm dyn}$, SSP parameters, {\it and} velocity dispersion simultaneously. This indeed removes the dependency on age and \afe, and interestingly implies a {\it negative} correlation of $\alpha_{\rm dyn}$ with [Z/H].

\vspace{-0.5cm}
\section{Conclusions}

The IMF normalisation inferred via stellar dynamics appears to show weaker correlations with the stellar population parameters than studies measuring the IMF directly from stellar light. Moreover, accounting for the dependence of stellar population parameters with velocity dispersion removes the apparent trends with age and \afe, while introducing a trend with [Z/H]. Resolving the differences of IMF inferences from independent methods remains a challenge for future work in this field.

\vspace{-0.5cm}
\section*{Acknowledgements}
\noindent
It is a great pleasure to thank the organisers and attendees for an exceptionally enjoyable conference, and to my colleagues in the \atlas\/ team for making this work possible.

\end{document}